\pdfoutput=1

\documentclass[sigconf]{acmart}

\usepackage{booktabs} 
\usepackage{algorithm}
\usepackage{algorithmicx}
\usepackage{algpseudocode}
\usepackage{amsmath}
\usepackage{amssymb}
\usepackage{flexisym}
\usepackage{makecell}
\usepackage{tikz}
\newcommand*\circled[1]{\tikz[baseline=(char.base)]{
            \node[shape=circle,draw,inner sep=2pt] (char) {#1};}}
\setcopyright{rightsretained}

\acmConference[FPGA'19]{The ACM/SIGDA International Symposium on Field-Programmable Gate Arrays}{2019}{Seaside, California USA}
\acmYear{2019}
\copyrightyear{2019}

\begin{document}
\title{Sparse Winograd Convolutional neural networks on small-scale systolic arrays}

%
%
%
%

\author{Feng Shi, Haochen Li, Yuhe Gao, Benjamin Kuschner, Song-Chun Zhu}
\affiliation{
  \institution{University of California Los Angeles}
  \streetaddress{UCLA Computer Science Department, 404 Westwood Plaza, Engineering VI}
  \city{Los Angeles}
  \state{USA}
  \postcode{90095-1596}
}
\email{{shi.feng, sczh}@cs.ucla.edu}
\renewcommand{\shortauthors}{F. Shi et al.}

\begin{abstract}

The reconfigurability, energy-efficiency, and massive parallelism on FPGAs make them one of the best choices for implementing efficient deep learning accelerators. However, state-of-art implementations seldom consider the balance between high throughput of computation power and the ability of the memory subsystem to support it. In this paper, we implement an accelerator on FPGA by combining the sparse Winograd convolution, clusters of small-scale systolic arrays, and a tailored memory layout design. We also provide an analytical model analysis for the general Winograd convolution algorithm as a design reference. Experimental results on VGG16 show that it achieves very high computational resource utilization, $20\times \sim 30\times$ energy efficiency, and more than $5\times$ speedup compared with the dense implementation.

\end{abstract}

%
%

%


\maketitle


\section{Introduction}
Convolutional neural network (CNN) is a class of deep learning algorithms which has become dominant in various computer vision tasks \cite{Zhang2018InterpretingCK, liu2016ssd}, 
so it is attracting research on acceleration for computational and power efficiencies.
The core computations in the algorithm are convolution operations with multi-dimensional data, e.g. 3-\textit{D} feature maps (FM) and 4-\textit{D} filters, which require a high density of memory accesses and high throughput of the computation engine. One research topic emerging in recent years is to deploy the convolution operations onto FPGAs \cite{Zhang2015, Suda2016, Zhang2016, 7929549}, since FPGAs consist of massive compute units, e.g. DSP blocks, and storage elements interconnected by reconfigurable switch blocks.
The most recent works on systolic array-based FPGA accelerators \cite{Wei2017, Cong2018} deliver significant performance improvement on the automation of high-level synthesis (HLS) design flow. Unlike the works \cite{LeCun2011, Wei2017}, which first construct 2-\textit{D} mesh architecture for systolic array then let the loops of codes to fit on these arrays (bitstream generated once), we recursively break the memory layout down to small blocks then map these blocks onto small-scale systolic arrays to perform multiplications of submatrices, and share these submatrices among working arrays to reduce required memory bandwidth.
Another performance improvement can be achieved from algorithmic perspective by applying the Winograd transform. This approach attracts more and more attention from researchers since its first GPU implmentation \cite{Lavin2016FastAF}. Winograd CNN accelerators on FPGAs are also well studied recently \cite{7929549, Aydonat2017}; however, the greater volume after the Winograd transformation is stressing on FPGAs. To handle this issue we adopt an efficient memory layout, adopt the pruned Winograd weights \cite{choi2018} and their elaborate hardware, and extend the computation into 3-\textit{D}. Pruning neural networks has been proven to greatly decrease both latency and energy consumption for all range of devices \cite{han2015deep_compression}. The major contributions are summarized in the following:
\begin{itemize}
\item \textbf{Unified small-scale systolic arrays for both Winograd transform and matrix multiplications}. We maximize the reusability of the existing design, e.g. RTL, for multiple modules. These modules share common characteristics, like matrix multiplication alike arithmetic operations.
\item \textbf{Efficient memory access layout.} We employ a recursive memory access pattern to increase locality of buffers. This pattern significantly impacts the overall performance.
\item \textbf{Block-based sparse matrix compression.} We employ this compression technique to adopt the above mentioned recursive memory layout.
\item \textbf{A comprehensive model analysis of Winograd convolution.} We propose an analytical model to investigate the performance and energy consumption, and based on the analysis we use the conclusion as our design guidance.
\end{itemize}


\section{Background}

\subsection{Spatial Convolution}
The convolution layer in a feedforward pass takes $C$ channels of $H \times W$ feature maps \textit{D} 
as input, and convolve each of $K$ filters of dimension $C \times r \times r$ with the input feature maps to produce $K$ output featre maps, \textit{Y}, 
of dimension $\left( H - r + 1 \right) \times \left( W - r + 1\right)$. Let $s$ be the stride and assume that the width and height of the filters are the same, then the mathematical description of the convolution is
\begin{equation} 
Y_{k, i, j} = \sum_{t=1}^C \sum_{p=1}^r \sum_{q=1}^r G_{k, t, p, q} \times D_{t, i * s + p, j * s + q}
\end{equation}

\subsection{Winograd Algorithm}
Winograd proposed an efficient algorithm for short convolutions \cite{1980-winograd} in computing of finite impulse response (FIR) filters in the signal processing field. \cite{Lavin2016FastAF} extends the Winograd algorithm to convolutional neural networks on GPU and CPU.

By applying Winograd transform to an r-tap FIR filter denoted as $F\left(m, r\right)$, which computes $m$ outputs with  the filter size of $r$, the number of multiplications is reduced from $m \times r$, if through the spatial convolution, to $m + r + 1$.

\subsubsection{1-D Winograd Convolution}

Taking $F\left(2, 3 \right)$ as an example, Winograd algorithm first transforms an input vector $d = \left(d_0, d_1, d_2, d_3\right)$ and filter $g = \left(g_0, g_1, g_2\right)$ into $j = \left(j_0, j_1, j_2, j_3\right)$ and $h = \left(h_0, h_1, h_2, h_3\right)$ respectively through


\begin{align*}
    j_0 = d_0 - d_2, & \quad h_0 = g_0 \\
    j_1 = d_1 + d_2, & \quad h_1 = \frac{g_0 + g_1 + g_2}{2} \\
    j_2 = d_2 - d_1, & \quad h_2 = \frac{g_0 - g_1 + g_2}{2} \\
    j_3 = d_1 - d_3, & \quad h_3 = g_2  \label{eq:wino_j_h}
\end{align*}
Next, element-wise multiplications are performed: 

\begin{equation} \label{eq:wino_elmt_wise}
  c_0 = j_0 \times h_0, \, c_1 = j_1 \times h_1, \,  c_2 = j_2 \times h_2, \, c_3 = j_3 \times h_3
\end{equation}
Finally, the output $y = \left(y_0, y_1\right)$ can be generated via:
\begin{equation} 
  y_0 = c_0 + c_1 + c_2, \quad y_1 = c_1 - c_2 - c_3
\end{equation}
The matrix form of the above procedure can be written as $y = A^T \left[ \left(G g \right) \odot \left(B^T d \right)\right] $, where $\odot$ represents element-wise multiplication and
\begin{align*}
    A^T =
      \begin{bmatrix}
      1 & 1 & 1 & 0 \\
      0 & 1 & -1 & -1
      \end{bmatrix}\,
    G = \begin{bmatrix}
        1 & 0 & 0 \\
        \frac{1}{2} & \frac{1}{2} & \frac{1}{2} \\
        \frac{1}{2} & -\frac{1}{2} & \frac{1}{2} \\
        0 & 0 & 1
        \end{bmatrix} \,
    B^T = \begin{bmatrix}
        1 & 0 & -1 & 0 \\
        0 & 1 & 1 & 0 \\
        0 & -1 & 1 & 0 \\
        0 & 1 & 0 & -1
        \end{bmatrix}
\end{align*}

The element-wise product in \eqref{eq:wino_elmt_wise} requires $m + r - 1 = 4$ multiplications, whereas the direct method does $m \times r = 2 \times 3 = 6$ multiplications.

\subsubsection{2-D Winograd Convolution} \label{subsec:2d_wino}
The 1-D Winograd algorithm can be easily extended to 2-D or higher dimensional convolutions by being nested with itself. 2-D Winograd algorithm $F\left(m \times m, r \times r\right)$ can be formulated as follows,
\begin{equation} \label{eq:2d_wino}
    Y = A^T \left[ \left(G g G^T\right) \odot \left(B^T d B\right) \right] A
\end{equation}
where $d$ and $g$ are tiles of input and the filter, having size of $l \times l$ ($l = m + r - 1$) 
and $r \times r$, respectively. The size of the output tile $Y$ is $m \times m$. \\
For larger input images, the Winograd transform is performed with the overlapping of tiles, with overlapping size $r - 1$, along each dimension. When applying Winograd algorithm to a convolution layer of CNNs, the tiles along the channel dimension of this layer can be fetched simultaneously and each of them is applied with \eqref{eq:2d_wino}.

\begin{figure}[!ht]
  \includegraphics[width=.44\textwidth]{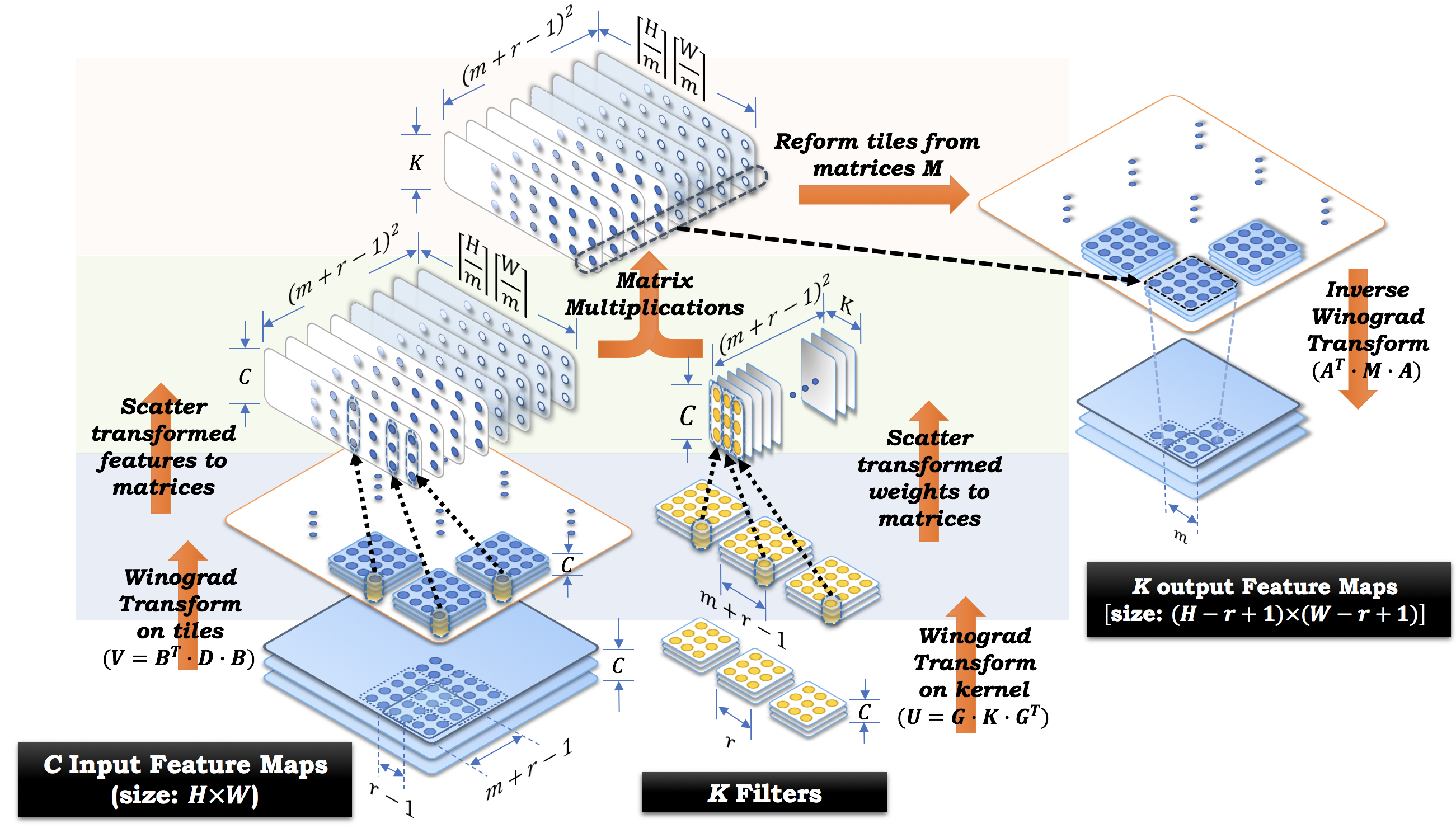}
  \caption{An overview of Winograd convolution layer.}
  \label{fig:wino_overview}
\end{figure}
\section{Algorithm and Optimizations}
This section gives an overview of our algorithm and presents several optimization methods. Fig. \ref{fig:wino_overview} shows the overview of our algorithm which consists of three stages of the Winograd-based convolution: input feature map and kernel transformations, matrix multiplications, and the inverse transformation of the output feature maps. These three stages form the pipeline of the data flow of our system design.
\subsection{Reduction to Matrix multiplication} \label{wino_matmul}
By reformulating \eqref{eq:2d_wino} with the augmentation on the channel dimension, filter $k$, tile coordinates $\left(\tilde{x}, \tilde{y}\right)$, and substitution of $U = G g G^T$ and $V = B^T d B$, we get
  \begin{equation}
    \label{eq:2d_wino_multi_chnl}
    Y_{k, \tilde{x}, \tilde{y}} = A^T \left[ \sum_{c=1}^C U_{k, c} \odot V_{c, \tilde{x}, \tilde{y}} \right] A
  \end{equation}
The summation part inside the parenthesis of \eqref{eq:2d_wino_multi_chnl} can be disentangled into $(m + r - 1)^2$ individual multiplication of a matrix of size $\left(C \times K \right)$ with another of size $\left(C \times \lceil H/m \rceil \lceil W/m \rceil \right)$.

\begin{align*}
    \mathcal{M}_{ k, \tilde{x}, \tilde{y} } = \sum_{c=1}^C U_{k, c} \odot V_{ c, \tilde{x}, \tilde{y} } \quad \xrightarrow[ \text{$\left(\tilde{i}, \tilde{j}\right)$ of tile} ]{\text{collapsing $\left(\tilde{x}, \tilde{y} \right)$ to $b$}} \nonumber \\
    \mathcal{M}_{\left(k, b\right)}^{\left(\tilde{i}, \tilde{j}\right)} = \sum_{c=1}^C U_{k, c}^{\left(\tilde{i}, \tilde{j}\right)} V_{c, b}^{\left(\tilde{i}, \tilde{j}\right)} \label{eq:wino_2_mtx}
\end{align*}


Another benefit of this reformation into matrix multiplications is that the number of inverse transforms has also been reduced over $C$ channels \cite{Lavin2016FastAF}, since the factorization of inverse transform along channels amortizes the cost. With this reformation, the matrix multiplications are then efficiently implemented on FPGAs.

\subsection{Matrix multiplications and memory access patterns}
\begin{figure}[!ht]
  \includegraphics[width=.44\textwidth]{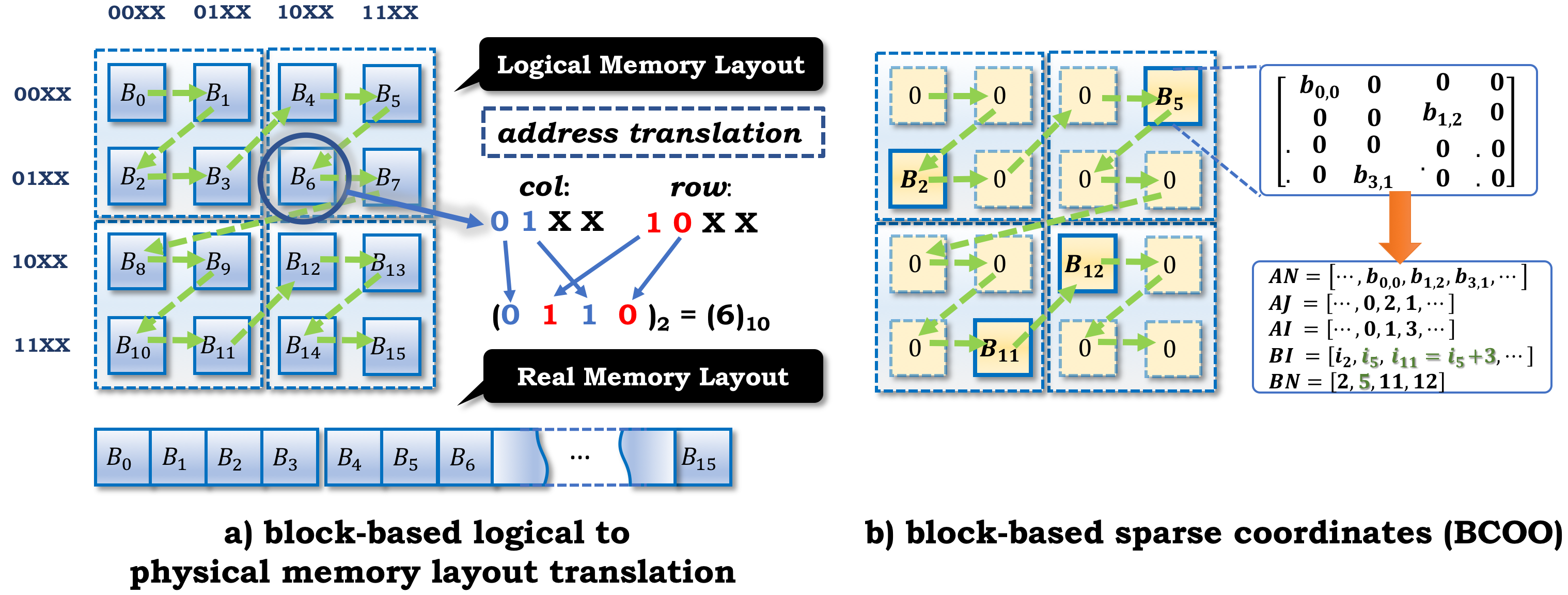}
  \caption{Z-Morton memory layout for both dense and sparse matrix \cite{Frigo_cache, DEEPA2012916}: $(a)$ the translation from logical layout to physical layout, $(b)$ the block-based compressed coordinates (BCOO, $l \times l$ block and $l=4$ for our design) for pruned Winograd weights}
  \label{fig:z_morton_addr}
\end{figure}
As described in section \ref{wino_matmul}, Winograd convolution can be computed efficiently with matrix multiplications on GPUs or FPGA platforms. To optimze the performance of matrix multiplication, we employ the Z-Morton memory layout \cite{Frigo_cache}, which has been widely studied for the Cache oblivious algorithms on multithreaded CPUs \cite{Frigo_cache, 1058095} and image processing on FPGAs \cite{DEEPA2012916}. This memory layout increases both spatial and temporal locality of memory accesses of matrix multiplication and arithmetic operations \cite{Frigo_cache}.
  \begin{algorithm}[H]
    \caption{Divide and Conquer Matrix Multiplication}\label{algo:matmul}
    \begin{algorithmic}[1]
      \Function{recursive-matmult}{$A, B, C$}
        \State $n = A.rows$
        \If{$n == l$} \Comment $l$ is the smallest tiling size
          \State $c_{1,1} = a_{1,1} \times b_{1, 1}$ \Comment matrix multiply of $l \times l$  tiles
        \Else
          \State partition $A$, $B$, and $C$ into tiles of size $\frac{n}{2} \times \frac{n}{2}$
          \State $C_{1,1} = \Call{recursive-matmult}{A_{1,1},B_{1,1}}$
          \State $\qquad + \Call{recursive-matmult}{A_{1,2},B_{2,1}}$
          \State $C_{1,2} = \Call{recursive-matmult}{A_{1,1},B_{1,2}}$
          \State $\qquad + \Call{recursive-matmult}{A_{1,2},B_{2,2}}$
          \State $C_{2,1} = \Call{recursive-matmult}{A_{2,1},B_{1,1}}$
          \State $\qquad + \Call{recursive-matmult}{A_{2,2},B_{2,1}}$
          \State $C_{2,2} = \Call{recursive-matmult}{A_{2,1},B_{1,2}}$
          \State $\qquad + \Call{recursive-matmult}{A_{2,2},B_{2,2}}$
        \EndIf
        \State ${\bf return} \, C$
      \EndFunction
    \end{algorithmic}
  \end{algorithm}

Z-Morton uses a \textit{divide and conquer} approach to access the memory as in Fig. \ref{fig:z_morton_addr} $\left(a\right)$. It is actually derived from the recursive matrix multiplication described in \textit{Algorithm} \ref{algo:matmul}. Compared with Strassen's algorithm, the latter is not cache-friendly in real situations, whereas the former can provide notable improvement in performance \cite{1058095}.
Note, instead of implementing the algorithm exactly, we unrolled memory access order to reorganize the memory layout.

The physical memory layout in FPGAs is essentially linear, Fig \ref{fig:z_morton_addr} ($a$) also provides an example of translating the logical block address to physical block address.
As shown in Fig. \ref{fig:z_morton_addr} ($a$), the address translation is easily implemented with LUTs in FPGAs by interleaving the bits of the logical column and row addresses to generate the physical address of a block.

\subsection{Pruned Winograd weights and memory access patterns}
After pruning the Winograd weights, we store them in a block-based sparse coordinates format (BCOO)\textendash only those $4 \times 4$ blocks containing nonzeros will be compressed and stored. Fig. \ref{fig:z_morton_addr} ($b$) shows an example where the block $B_5$ is a $4 \times 4$ tile, and it has 3 nonzeros. The information of these nonzeros are stored into vectors $BN$, $BI$, $AI$, $AJ$, and $AN$.
$BN$ contains the block number for each block in memory layout, e.g. 5 for $B_5$. $BI$ is the list of starting indices of each block within the other three arrays, e.g. $i_5$ of $BI$ refers to the starting index in $AI$, $AJ$, and $AN$ of information corresponding to $B_5$. Elements in $AI$ and $AJ$ represent the row and column number of the nonzeros in its own block, respectively, and $AN$ stores the value of the corresponding nonzero. For $B_5$, the values of nonzeros are $b_{0,0}$, $b_{1,2}$, and $b_{3,1}$, the corresponding column numbers are 0, 2, 1 and row numbers are 0, 1, 3 in $AJ$ and $AI$, respectively. The compressed blocks are still fetched following the order determined by Z-Morton layout.

\section{Architecture Design}
This section discusses our implementation of accelerator for Winograd convolution. The most time-consuming parts in the computation pipeline are the Winograd transform for feature maps and matrix multiplications. In our design, we propose using unified small-scale systolic arrays, of size $l \times l $ $\left( l = m + r - 1 \right)$, for both these arithmetic operations.

\subsection{Winograd transform by Systolic Arrays}
Recall the $2$-$D$ Winograd transform nesting 2 transform matrices, $B^T \cdot D \cdot B$.
\begin{figure}[!ht]
  \includegraphics[width=.40\textwidth]{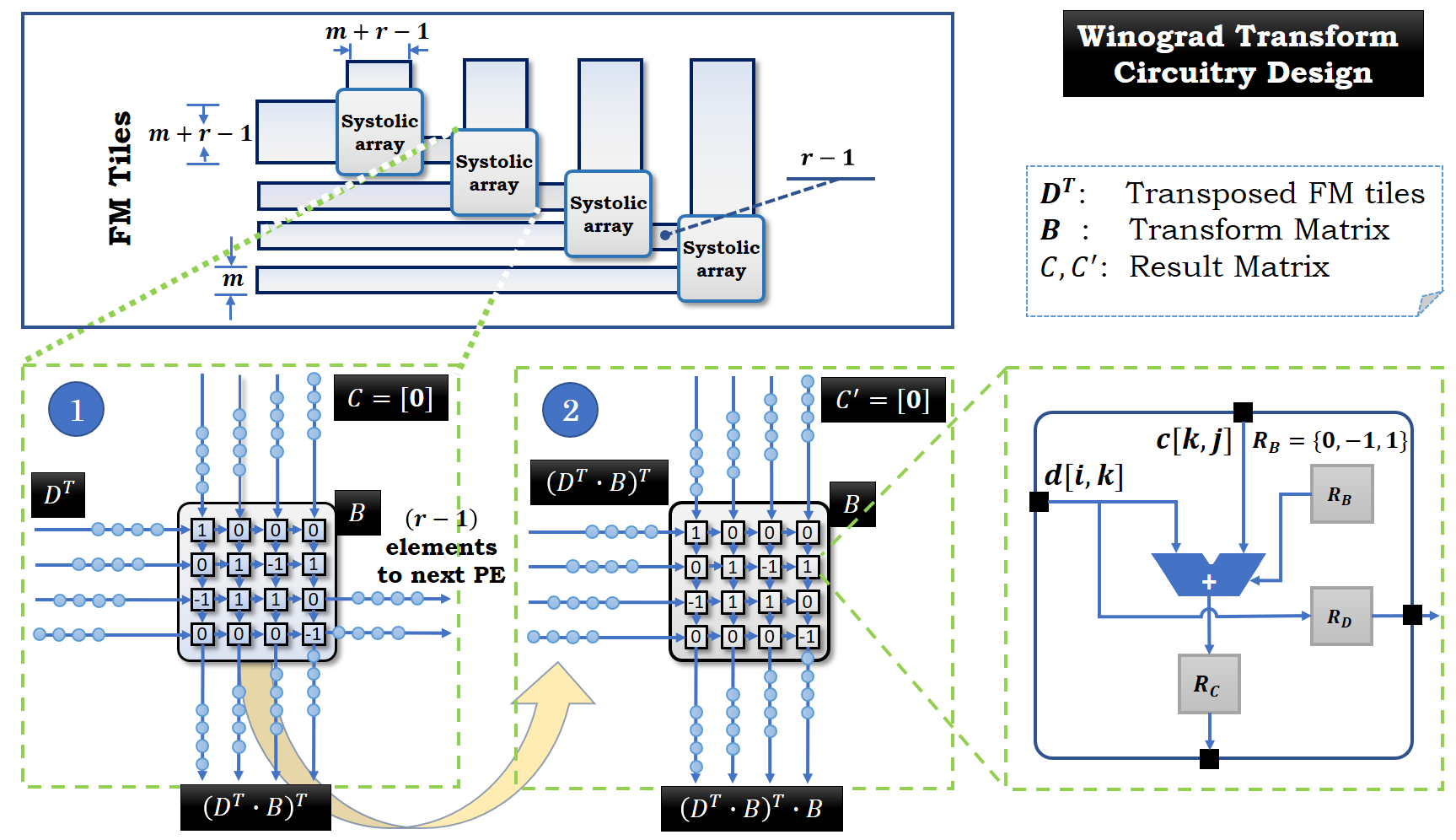}
  \caption{Small-scale Systolic Arrays for Winograd Transform}
  \label{fig:wt_systolic}
\end{figure}
Instead of directly computing $B^T \cdot D \cdot B$, we change it into $\left( D^T \cdot B \right)^T \cdot B$. Thus, we let transform matrix $B$ be stationary inside the systolic arrays. In the first iteration \circled{1} of the Fig. \ref{fig:wt_systolic} $D^T$ passes through systolic arrays to operate with \textit{B} and the output is $P = \left(C + D^T \cdot B\right)^T$ (no additional transpose needed). This intermediate result $\left( D^T \cdot B \right)^T$ feeds back to systolic arrays as "new  $D^T$" in the second iteration \circled{2}.
Then $P\textprime = C\textprime + P \cdot B = \left( D^T \cdot B \right)^T \cdot B = B^T \cdot D \cdot B$ is the final resutl. Note that \textit{C} and \textit{C$\textprime$} are zero-matrices and there is no multiplication occured inside these systolic arrays\textendash the value of elements of \textit{B} is just used to control the adder\textendash such as, "1" for addition, "-1" for subtraction, and "0" for passing by the data to next processing element (PE) inside its systolic array.

The data sharing is through the overlapping of tiles, which has been described in section \ref{subsec:2d_wino}. Fig. \ref{fig:wt_systolic} illustrates that $\left( m + r - 1 \right)$ wide data stream into each systolic array, and among these data, $\left(r - 1\right)$ of them travel through the current systolic array and are forwarded to the next systolic array at the same direction. The output is streamed out in the orthogonal direction after two iterations as stated previously, and is transfered into shift-registers for scattering into matrices.

\subsection{Matrix Multiplication by Systolic Arrays}
\begin{figure} 
  \center
  \includegraphics[width=.45\textwidth]{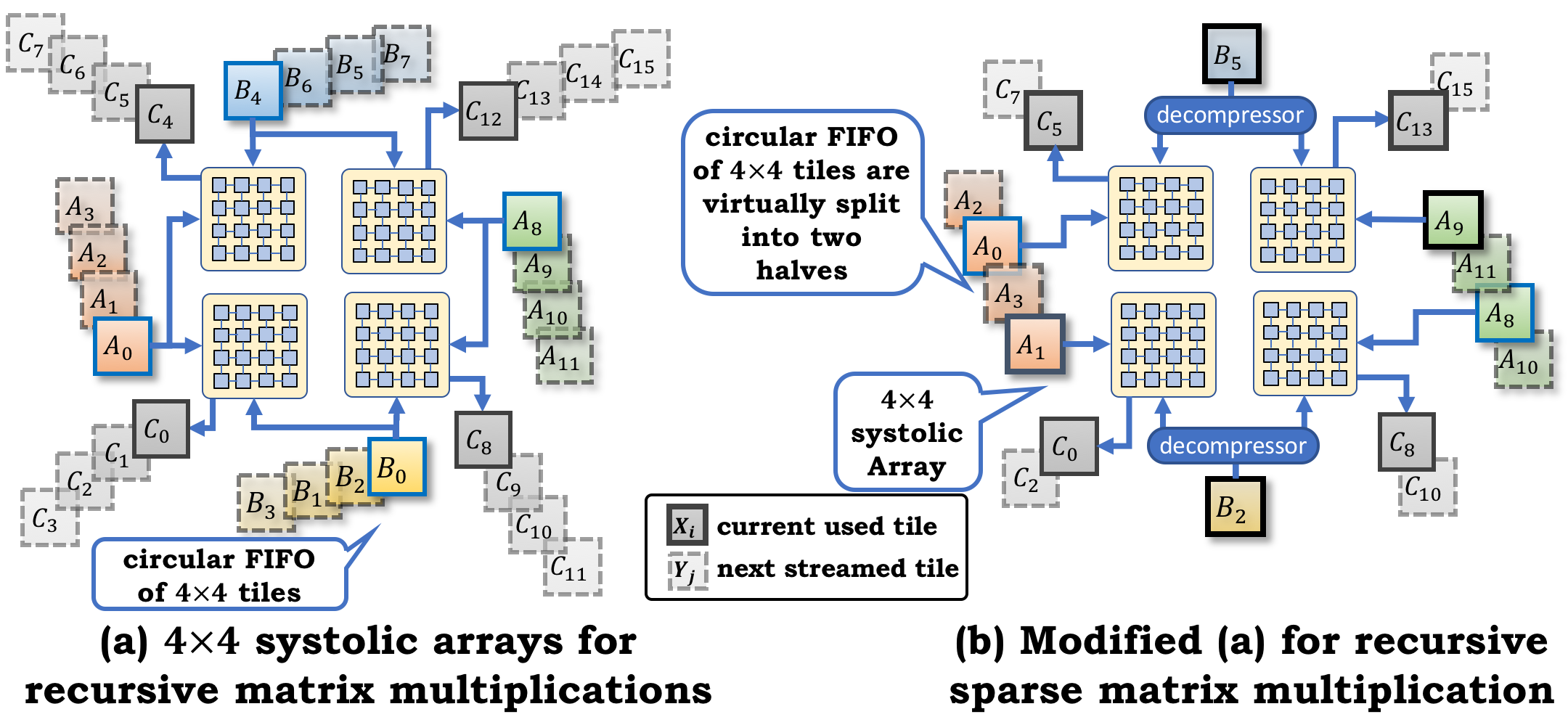}
  \caption{Systolic Arrays for \textit{Algorithm \ref{algo:matmul}}: $(a)$ the original design for dense case, $(b)$ modified architecture for sparse case}
  \label{fig:mm_systolic}
\end{figure}

To perform the recursive matrix multiplication \textit{Algorithm \ref{algo:matmul}} with hardware, we conceive the cluster of small-scale systolic arrays. Each cluster consists of 4 $l \times l$ systolic arrays ($l = 4$ for our case) and a set of shared circular FIFO built by shift-registers, shown in Fig. \ref{fig:mm_systolic}.
To understand how this cluster works, let us examine the example from Fig. \ref{fig:z_morton_addr}. By unrolling the recursive code given by \textit{Algorithm \ref{algo:matmul}} and using the tiles of matrices organized by Z-Morton layout, we calculate sub-matrix $C_{0}$ by summing up the products of submatrices $A_0 \times B_0$ and $A_1 \times B_2$, $C_4$ by sum of $A_0 \times B_4$ and $A_1 \times B_6$, and so on.
{\small
  \begin{align*}
      C_0 \, &\text{+=} \, A_0 \times B_0 + A_1 \times B_2;\\
      C_4 \, &\text{+=} \, A_0 \times B_4 + A_1 \times B_6;\\
      C_8 \, &\text{+=} \, A_8 \times B_0 + A_9 \times B_2;\\
      C_{12} \, &\text{+=} \, A_8 \times B_4 + A_9 \times B_6; \\
      \cdots \\
      C_0 \, &\text{+=} \, A_4 \times B_8 + A_5 \times B_{10};\\
      C_4 \, &\text{+=} \, A_4 \times B_{12} + A_{5} \times B_{14};\\
      \cdots
  \end{align*}
}

As shown in Fig.\ref{fig:mm_systolic} (\textit{a}), $A_0$ is shared by northwest and southwest systolic arrays, $A_8$ is shared by northeast and southeast systolic arrays, and so on. After the first iteration, the partial results of $C_0$, $C_4$, $C_8$, and $C_{12}$ are produced and stored inside the corresponding systolic arrays. In the second iteration, the blocks $A_1$, $A_9$, $B_4$, and $B_9$ get into their corresponding systolic arrays and perform the matrix multiplications, and their products are accumulated to the partial results, which still stay in their systolic arrays from iteration 1. At iteration 3 the results of $C_0$, $C_4$, $C_8$, and $C_{12}$ are spilled out, and systolic arrays continue to work on the partial results of $C_1$, $C_5$, $C_9$, and $C_{13}$. This procedure continues until all the submatrices are calculated. Also the sharing of circular FIFOs reduces the memory bandwidth requirement by 4 folds.

When the computation is comprised of sparse matrix multiplications, we need some modifications on the cluster of systolic arrays. First, each of the circular FIFOs which supply the compressed Winograd weight blocks need to be equipped with a decompressor. Second, the circular FIFOs for Winograd feature maps are virtually split into two halves since some Winograd feature maps blocks are no longer shared between the systolic arrays. The overall memory access pattern is now determined by how the sparse blocks distributed in the memory layout. Take the sparse blocks $B_2$ and $B_5$ from Fig. \ref{fig:z_morton_addr} for example; now we notice that the computation of $C_0$ becomes $A_1 \times B_2$ only, $C_8$ becomes $A_9 \times B_2$, block $B_2$ is still shared by the products of submatrices $C_0$ and $C_8$.

\subsection{Extends the computation into third dimension}
\begin{figure}[h!]
  \center
  \includegraphics[width=.34\textwidth]{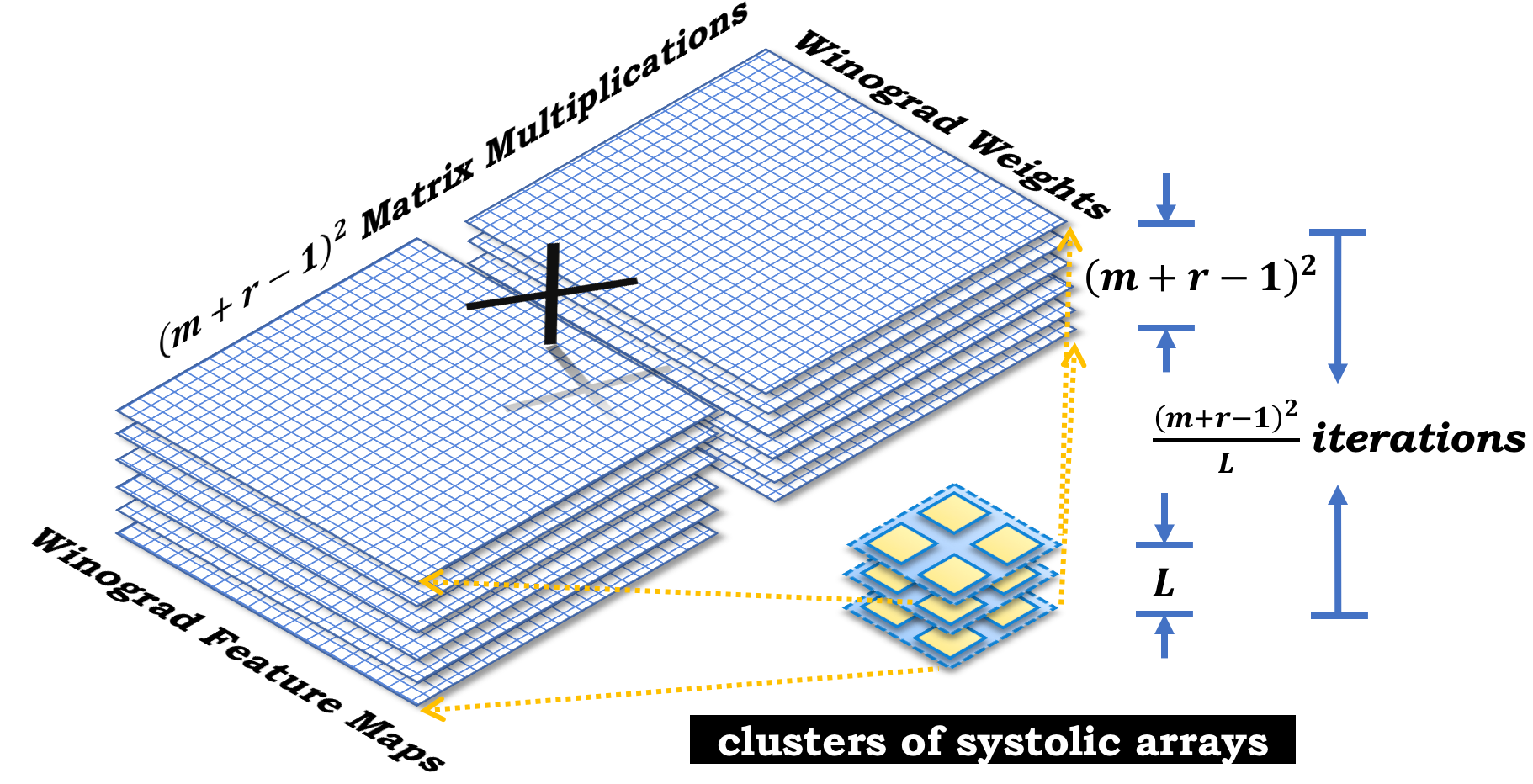}
  \caption{Extension of computation to 3-$D$ dimension}
  \label{fig:3d_systolic}
\end{figure}
Whenever the computation resource is available, we can extend the computation into higher dimensions. As we have analyzed in section \ref{wino_matmul}, there are $\left(m + r - 1 \right)^2$ independent matrix multiplications, and they can be executed in parallel with several clusters of systolic arrays as demonstrated in Fig. \ref{fig:3d_systolic}. With this enhencement, the DSP utilization and throughput of the FPGA system are dramatically improved. In our design, we organize the DSPs into 8 clusters due to the limited amount of DSPs in our FPGA board.

\subsection{Extension to other types of layers}
In addition to convolution layers, fully-connected (FC) layers are essentially computed through matrix multiplications. Therefore, the techniques previously discussed can be also employed to FC layers.
ReLU layers and Max Pooling layers are easily implemented by accompanying comparators to the output buffers.

\section{Design Space Exploration}
\subsection{Model Analysis}
A detailed study of the complexity of Winograd convolution is conducted in the following subsections, it helps us to design an optimzed accelerator for both dense and sparse cases.

\subsubsection{Data Layout of Winograd transform}
As previously mentioned, the input feature maps are fed in system in real-time. It's not convenient to prune them during the inference, and it will increase the difficulty in system design. Moreover, the multiplication of a sparse matrix with a dense one does not necessarily produce another sparse matrix. In such case, our analysis keeps the same characteristics of feature maps for both dense and sparse cases.
The volume of $i^{th}$ Winograd convolution layer $D_{wi}^{i}$, the volume of corresponding Winograd weights $D_{wk}^{i}$ (without pruning), and the volume of the results $D_{wo}^{i}$ before the inverse Winograd transform can be computed as
\begin{equation} \label{eq:nb_wi}
    D_{wi}^{i} = \left\lceil \frac{H}{m} \right\rceil \times \left\lceil \frac{W}{m} \right\rceil \times C \times l^2 \approx \left(\frac{l}{m}\right)^2 \times H \times W \times C
\end{equation}
\begin{equation} \label{eq:nb_wo}
    D_{wo}^{i} = \left\lceil \frac{H}{m} \right\rceil \times \left\lceil \frac{W}{m} \right\rceil \times K \times l^2 \approx \left(\frac{l}{m}\right)^2 \times H \times W \times K
\end{equation}
\begin{equation} \label{eq:nb_wk}
    D_{wk}^{i} = C \times K \times l^2
\end{equation}
The Winograd transform dilates both the input feature maps and weights by a scale factor of $\left(\frac{l}{m}\right)^2$, e.g. when $m$ takes value of 2 and $r$ of 3, the transformed feature maps and weights require roughly 1.78 times larger storage. The increased volume of the storage not only affects the latency of computations due to the drastically slow access speed, but also causes more energy consumption.


\subsubsection{Arithmetic complexity}
The arithmetic complexity greatly depends on the data layout since the volume of feature maps and weights decides how much data does the algorithm needs to process.
The number of multiplications performed by Winograd convolution layer $i$ is
   \begin{align*}
      M_W^i = \left\lceil \frac{H}{m} \right\rceil \cdot \left\lceil \frac{W}{m} \right\rceil \cdot C \cdot K \cdot l^2
      \approx H \cdot W \cdot C \cdot K \cdot \left(\frac{l}{m}\right)^2
  \end{align*}

The number of additions involved in matrix multiplications is

   \begin{align*}
      S_W^i = \left\lceil \frac{H}{m} \right\rceil \cdot \left\lceil \frac{W}{m} \right\rceil \cdot \left(C - 1\right) \cdot K \cdot l^2 
      \approx H \cdot W \cdot \left(C - 1\right) \cdot K \cdot \left(\frac{l}{m}\right)^2
  \end{align*}
The number of additions required by Winograd transforms are $S_B$ and $S_A$ for $\left(B^TdB\right)$ and $\left(A^T \left[\mathcal{M}_{k, \tilde{x}, \tilde{y}}\right] A\right)$ respectively. In most cases, Winograd transform matrices $B$ and $A$ are sparse, therefore, \eqref{eq:nb_add_wb} and \eqref{eq:nb_add_wa} utilize the operator $nnz\left( \cdot \right)$ (number of nonzeros).
  \begin{equation} \label{eq:nb_add_wb}
         S_B^i = 2 \times \left\lceil \frac{H}{m} \right\rceil \times \left\lceil \frac{W}{m} \right\rceil \times C \times K \times l \times \left[nnz\left(B\right) - l \right]
  \end{equation}
  \begin{equation} \label{eq:nb_add_wa}
     S_A^i = 2 \times \left\lceil \frac{H}{m} \right\rceil \times \left\lceil \frac{W}{m} \right\rceil \times C \times K \times l \times \left[nnz\left(A\right) - m \right]
  \end{equation}
The Winograd weights are pre-calculated and stored in memory, so the overhead of computing Winograd weights has not been taken into account.

\subsubsection{Optimal Winograd transform and the corresponding "$m$" }

\begin{figure}[h!]
  \includegraphics[width=.37\textwidth]{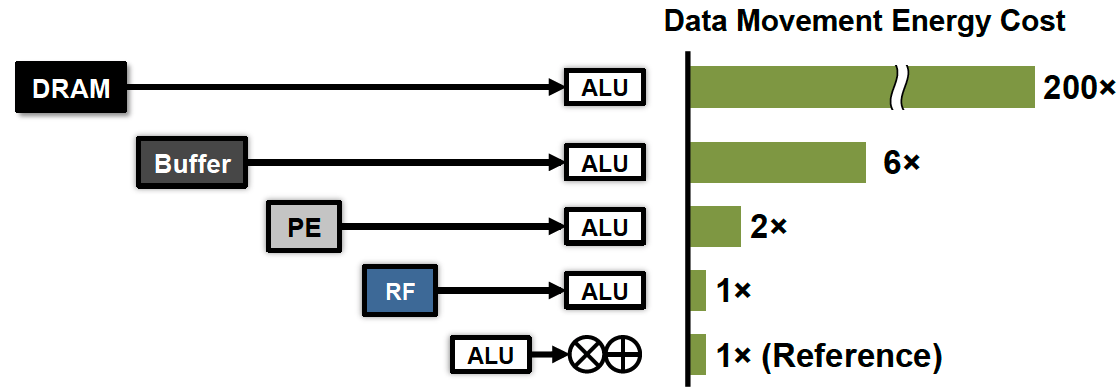}
  \caption{Data movement energy comparison among memory hierarchies \cite{Sze2017HardwareFM}}
  \label{fig:mem_hier_energy}
\end{figure}
When the value of $r$ is specified, e.g. $r=3$ for every layer of $VGG$, the value of $m$ is crucial for determining both the power consumption and the arithmetic complexity. Furthermore, the calculation of the optimal power consumption is straightforward, whereas the optimal computation time is much more complicated to evaluate. Since the degree of parallelism and the memory access patterns are dynamic, these uncertain factors hinder accurate estimation of optimal computation time in an obvious mathematical analysis. Therefore, we focus on the analysis of achieving the optimal power consumption as the reference.

As shown in Fig. \ref{fig:mem_hier_energy}, the energy consumption for local (e.g. buffers, FIFOs) and external memory accesses are several times and orders of magnitude higher than arithmetic operations, respectively \cite{Sze2017HardwareFM}. Let us assume for the sake of simplicity that every storage element in both local and external memory is accessed exactly once, transformed feature maps are stored in local memory after Winograd transform, and the Winograd weights are read from external memory.

Let $E_{me}$ and $E_{ml}$ be the unit energies consumed by an access to the external memory and an access to the local memory, respectively. Let $E_{mul}$ and $E_{add}$ be the unit energies consumed by a multiplication operation and an addition operation, respectively. Then the total energy consumption of layer $i$ is
\begin{align*}
      E_{tot}^i = E_{ml} \cdot \left(D_{wi}^i + D_{wo}^i\right) + E_{me} \cdot D_{wk}^i + \\
                  E_{mul} \cdot M_W^i + E_{add} \cdot \left(S_W^i + S_B^i + S_A^i\right)
\end{align*}
Another fact derived by eq. \eqref{eq:nb_wi} and \eqref{eq:nb_wk} is that greater $m$ generates less elements of the transformed feature maps but more elements of the transformed weights. This fact indicates that the pruning of Winograd weights is more efficient with greater $m$.

After having given the above formulas and summarizations, we conduct the analysis and experiments in section \ref{sec:experiment_energy}.


\begin{table}[h!]
    \caption{number of parameters in each convolution layer of different stages in $VGG$ \cite{Simonyan14c} after Winograd transform (\textit{m}=2)}
    \label{tab:nb_params}
    \begin{tabular}{ccc}
      \toprule
      Stage \cite{Simonyan14c} & \makecell{\# of Winograd neurons} & \makecell{\# of Winograd weights} \\
      \midrule
      Conv$1\,\left(\times 2\right)$ & 12,845,056 & 65,536 \\
      Conv$2\,\left(\times 3\right)$ & 6,422,528	& 262,144 \\
      Conv$3\,\left(\times 4\right)$ & 3,211,264	& 1,048,576 \\
      Conv$4\,\left(\times 4\right)$ & 1,605,632	& 4,194,304 \\
      Conv$5\,\left(\times 4\right)$ & 401,408	& 4,194,304 \\
      Conv6 & 131,072	& 4,194,304 \\
      \bottomrule
   \end{tabular}
\end{table}
\section{Experimental Evaluation}
$VGG$ \cite{Simonyan14c} is one of the most popular and mature deep learning models which has been widely used in research and industry. In this work, we use $VGG16$ for our analysis and experiments.
\subsection{Experiment Setup}
For the CNN model part, we set the input feature map size to $224 \times 224 \times 3$, which are standard input dimensions for VGG pipeline.

Table \ref{tab:nb_params} shows the number of neurons and weights of each layer in different stages after the Winograd transform.
For the hardware part, we evaluate our design on an FPGA board, Xilinx Virtex Ultrascale XCVU095. Although it is not fabricated with the lastest technologies, and equips only with a medium amount of DSPs (768 DSPs), this configuration reveals better the performance gain than the lastest FPGAs since optimizations for FPGAs with scarce computation power is more representative.

  \begin{table*} 
    \caption{Comparison with State-of-the-art implementations}
    \label{tab:fpga_comp}
    \begin{tabular}{@{}cccccccc@{}}
    \toprule
    Impl. & FPGA'15 \cite{Zhang2015} & FPGA'16 \cite{Zhang2016} & FPGA'16 \cite{Suda2016}       & \multicolumn{2}{c}{DAC '17 \cite{Wei2017}}                       & \multicolumn{2}{c}{our impl.}                                                                                                                \\ \midrule
    FPGA                        & V7 VX485T    & Xilinx VC709 & Stratix-V GSD8 & \multicolumn{2}{c}{Arria10 GT1150}                & \multicolumn{2}{c}{V-Ultra XCVU095}                                                                                                          \\
    Precision                   & 32 bit float & 16 bit fixed & 8-16 bit fixed & 32 bit float & \multicolumn{1}{l}{8-16 bit fixed} & \multicolumn{2}{c}{8-16 bit fixed}                                                                                                           \\
    Frequency (MHz)             & 100          & 200          & 120            & 221.65       & 231.85                             & \multicolumn{2}{c}{150}                                                                                                                      \\
    Throughput (Gops/s)         & 61.6         & 354          & 47.5           & 460.5        & 1171.3                             & \begin{tabular}[c]{@{}c@{}}460.8/230.4 \\ (8 bit/16 bit fixed)\end{tabular} & \begin{tabular}[c]{@{}c@{}}921.6 (projected, \\ 8 bit fixed sparse)\end{tabular} \\
    DSP utilization             & 1120/1400    & 2833/3632    & 727/1963       & 1340/1523    & 1500/3046                          & \multicolumn{2}{c}{(512+256)/768}                                                                                                            \\
    Power efficiency (Gops/s/W) & 3.31         & 14.22        & 1.84           & \multicolumn{2}{c}{25.78}                         & \multicolumn{2}{c}{55.9}                                                                                                                     \\ \bottomrule
    \end{tabular}
  \end{table*}

\begin{figure}[h!]
  \includegraphics[width=.48\textwidth]{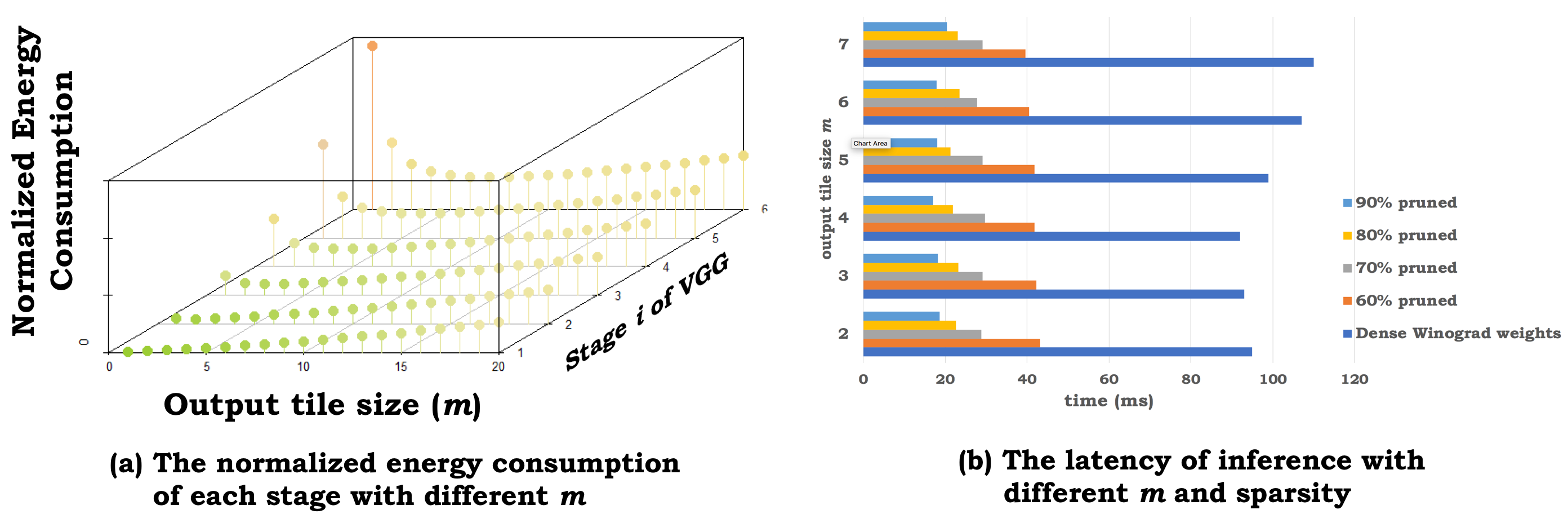}
  \caption{Energy consumption estimation and latency of Winograd convolution}
  \label{fig:energy_wino}
\end{figure}

\subsection{Experiment on energy consumption analysis} \label{sec:experiment_energy}

Fig. \ref{fig:energy_wino} (\textit{a}) plots the trend when different $m$ is applied. The simulations run by synthesis tools show that the design with small values of $m$ normally consume less energy. In order to simplify our design, we decide to use $m=2$, which eventually affects the dimension of our systolic arrays, tiling size, memory access patterns of our accelerator design, and so on. Although the plot indicates that $m=4$ might be the optimal value for the energy consumption, we are limited by other hardware resources in our FPGA system, but the situation might be different if designing with a different FPGA system. In Fig. \ref{fig:energy_wino} (\textit{b}) we provide the latencies for the inference by VGG with different configuration of \textit{m} and sparsity ranging from $60\%$ to $90\%$. For the best case, we achieve almost $5\times$ speedup.

 \begin{table}[h]
    \caption{Resource usage}
    \label{tab:fpga_rsc}
    \begin{tabular}{ccccc}
      \toprule
      Resources & LUTs & FF & BRAM & DSP \\
      \midrule
      Used & 241,202	& 634,136 & 1,480 & \makecell{512 (arith.) + 256 (wino.)} \\
      Available\cite{xilinx2015} & 537,600	& 1,057,200 & 1,728 & 768 \\
      Percentage & 44.9\%	& 60.8\% & 85.6\% & \makecell{67\% + 33\% =100\%} \\
      \bottomrule
    \end{tabular}
  \end{table}

\subsection{Results and analysis}

With $m=2$, we get the synthesized result with the resource usage as shown in Table \ref{tab:fpga_rsc}. The end-to-end comparison with the state-of-art CNN FPPGA accelerators is listed in Table \ref{tab:fpga_comp}. We achieve the highest DSP usage and power efficiency. Due to time limitations, we only test our design on a medium scale FPGA. In current design, we use four $4 \times 4$ systolic arrays as one cluster for one matrix multiplication, and stack 8 such clusters for eight matrix multiplications in parallel. Meanwhile, 16 $4 \times 4 $ systolic arrays work on the Winograd transform. In total, all 768 PEs are used. We will try to transfer our design to the latest and most powerful FPGA board in the future, and the performance will be improved further.

\section{Conclusion}
In this paper we propose a design with highly efficient recursive memory access layout for both dense and sparse Winograd convolutions, unified systolic arrays for both Winograd transforms and matrix multiplications, and a three dimensional compute engine for Winograd convolution. We also provide a comprehensive algorithmic level analysis for the performance model of the Winograd convolution. We achieve high computation power usage and high power efficiency in our design. There are several aspects that we can investigate further in the future. In particular, the automation design flow will help a lot to reduce the burden of development. And, the progress in memory technology is also a promissing solution as more and more new FPGA architecture incorporate such kind of brilliant concept.

%
%
\bibliographystyle{ACM-Reference-Format}
\bibliography{bibliography}

\end{document}